\documentclass[12pt,preprint]{aastex}
\begin{document}
 
\def\today{\number\year\space \ifcase\month\or  January\or February\or
        March\or April\or May\or June\or July\or August\or
September\or
        October\or November\or December\fi\space \number\day}
\def\fraction#1/#2{\leavevmode\kern.1em
 \raise.5ex\hbox{\the\scriptfont0 #1}\kern-.1em
 /\kern-.15em\lower.25ex\hbox{\the\scriptfont0 #2}}
\def\spose#1{\hbox to 0pt{#1\hss}}
\def\simlt{\mathrel{\spose{\lower 3pt\hbox{$\mathchar''218$}}
     \raise 2.0pt\hbox{$\mathchar''13C$}}}
\def\simgt{\mathrel{\spose{\lower 3pt\hbox{$\mathchar''218$}}
     \raise 2.0pt\hbox{$\mathchar''13E$}}}
\def\etal{et al. }

\title{On the Location of the Snow Line in a Protoplanetary Disk}
\author{M. Lecar\altaffilmark{1}, M. Podolak\altaffilmark{2}, D. Sasselov\altaffilmark{1} \& E. Chiang\altaffilmark{3,4}}
\altaffiltext{1}{Harvard-Smithsonian Center for Astrophysics, 60
Garden St., Cambridge, MA 02138}
\altaffiltext{2} {Dept. of Geophysics \& Planetary Sci., Tel Aviv University,
Tel Aviv, Israel}
\altaffiltext{3}{Astronomy Department, UC Berkeley, Berkeley, CA 94720}
\altaffiltext{4}{Alfred P.~Sloan Research Fellow}

\begin{abstract}

In a protoplanetary disk, the inner edge of the region where the
temperature falls below the condensation temperature of water is
referred to as the 'snow line'. Outside the snow line,
water ice increases the surface density of
solids by a factor of 4. The mass of the fastest growing planetesimal (the
'isolation mass') scales as the surface density to the 3/2 power.
It is thought that ice-enhanced surface densities
are required to make the cores of the
gas giants (Jupiter and Saturn)
before the disk gas dissipates. 
Observations of the Solar System's asteroid belt suggest that the
snow line occurred near 2.7 AU.
In this paper we revisit the theoretical
determination of the snow line. In a minimum-mass disk
characterized by conventional opacities and a mass
accretion rate of $10^{-8} M_{\odot}/{\rm yr}$, the snow line
lies at 1.6--1.8 AU, just past the orbit of Mars.
The minimum-mass disk, with a mass of 0.02~$M_{\odot}$, has a life time of
2 million years with the assumed accretion rate.
Moving the snow line past 2.7 AU requires that we increase the disk
opacity, accretion rate, and/or disk mass by factors ranging up to an order
of magnitude above our assumed baseline values.
\end{abstract}


\section{Introduction}
Most of the extra-solar planets that have been detected so far are
Jupiter-like gas giants. The most widely accepted theory for the formation
of gas giants is the 'core accretion' model (e.g., Pollack \etal 1996;
Rafikov 2004),
which requires a core of $5-15~M_{\oplus}$ (Guillot 2005).
Some of the extra-solar planets seem to have masses in this
range (McArthur et al.~2004;
Santos et al.~2004); they would
serve as cores if there were gas for them to accrete.

In the minimum-mass solar nebula (MMSN), the surface density of refractory
materials is about
$0.64~{\rm g}/{\rm cm}^2$ at 5~AU.
The surface density of all condensible material increases to
$2.7~{\rm g}/{\rm cm}^2$ once the volatiles (ices) freeze out.
The 'isolation masses' of the early
planetary embryos, after they have swept up all the material in their annular
feeding zones in the parent disk, is proportional to the $\frac{3}{2}$-th power
of the surface density. Taking the radial width of the feeding zone
to be $2{\sqrt{3}}$ Hill radii (Gladman 1993),
and using the ice-enhanced surface density,
we find that the isolation mass at Jupiter's distance
is about the mass of the Earth.
These embryos then merge to form the $5-15~M_{\oplus}$
cores of the gas giants. It has been traditionally believed that the
surface density needs to be enhanced by ices to
form the cores of the giant planets before disk gas dissipates.

Ice forms at (and beyond) the
'snow line' where the temperature falls below 145--170~K, depending on the
partial pressure of nebular water vapor.
Previous work (Hayashi 1981; Sasselov \& Lecar 2000)
neglected the dependence of the sublimation temperature on the gas
density. Podolak and Zucker (2004) showed that for the densities in the
MMSN, the sublimation temperature can be as low as 145~K.

In an earlier paper (Sasselov \& Lecar 2000) we
found the midplane
temperature dropped to 170~K at a distance of $1.5~{\rm AU}$
(the heliocentric distance of Mars) for a disk heated purely
by incident starlight (a ``passive'' disk that does not accrete).
The intent of that paper was to see if
close-in extra-solar planets could be formed {\em in situ}, i.e., if cores
weighing a few Earth masses could be formed at the distance of Mercury.
We were content to show that they could not. However, in our Solar System,
the snow line was definitely exterior to the orbit of Mars. The evidence
points to about $2.7~{\rm AU}$, in the outer asteroid belt
where icy C-class asteroids abound (Abe et al. 2000;
Morbidelli et al. 2000; Rivkin et al.~2002). Comets
are more water-rich by about a factor of 4, while the inner asteroid
belt is largely devoid of water. While this evidence has been questioned,
and alternatives proposed (e.g. Grimm \& McSween 1993), it appears that the solar
nebula at the time of planetesimal formation was hotter than the models
discussed by Sasselov \& Lecar (2000).

In this paper we revisit the issue of the snow line. We aim to find out how
global disk parameters (surface density, mass accretion rate, opacity)
affect the location where the snow transition occurs. We are concerned
with large-scale disk properties and ignore here local perturbations
due to protoplanets discussed by Jang-Condell \& Sasselov (2004). In Section 2
we describe our model for protoplanetary disks, and in Section 3
we calculate the temperature and density-dependent rates of
ice sublimation and condensation.
We present our results and conclusions in section 4.

\section{The Model} 
Our model is that of a disk that is heated not only by
steady mass accretion at rate $\dot{M}$,
but also by absorption of light emitted by the central star.
We work with a disk whose surface density is that of the MMSN:
${\Sigma}(r)={\Sigma}_0 (r/{\rm AU})^{-3/2}$, where $r$ is the disk radius,
$\Sigma$ is the total surface density in gas and condensibles,
and ${\Sigma}_0=1700~{\rm g}/{\rm cm}^2$.
We avoid explicitly accounting for the usual
dimensionless viscosity parameter, $\alpha$ (e.g., Frank, King, \& Raine 1992),
by fixing the value of $\dot{M}$ and using our prescribed
surface density law. These choices define an $\alpha$ that
is not constant with radius.

To estimate the midplane temperature, we first neglect
absorption of starlight and consider accretional heating only.
The flux emitted by a disk that steadily accretes mass
at rate $\dot{M}$ in a potential due to a central star of mass $M_{\star}$
and radius $R_{\star}$ is (Lynden-Bell \& Pringle 1974):

\begin{equation}
F_{\rm acc}(r) \equiv {\sigma}T_{\rm eff}^4 = \frac{3}{8\pi} \frac{GM_{\star}\dot{M}}{r^3} \left( 1-\sqrt{\frac{R_{\star}}{r}} \right).
\end{equation}

 Here $T_{\rm eff}(r)$ is the effective temperature corresponding
to the total flux released by accretional heating.
We use this effective temperature to
evaluate the midplane temperature of the disk
under the assumption that the accretional energy is transported
radiatively from the midplane to the surface.
Treating radiative diffusion in an optically thick
medium, we can safely adopt the Eddington approximation. We employ the
Rosseland optical depth,
$$
{\tau}_R = \int \limits_0^{\infty} \kappa {\rho}(r,z) dz,
$$
to derive the midplane temperature due to accretional heating only,
\begin{equation}
T_{\rm mid,acc}^4 = \frac{3}{4} \left( {\tau}_R +\frac{2}{3} \right) T_{\rm eff}^4.
\end{equation}
 Here $\rho(r,z)$ is the total mass density at radius $r$ and vertical
height $z$ above the midplane, and $\kappa$ is the Rosseland opacity.
The latter quantity is taken from D'Alessio, Calvet \& Hartmann (2001);
it is dominated by particle condensates and is a function of temperature.
It is uncertain insofar as the properties of the condensates---their mineral
composition, allotropic state, and distribution with size---are uncertain.
We make use of the dependence of the opacity on whether the temperature is
above (300~K) or below (100~K) ice sublimation as given by D'Alessio, Calvet \& Hartmann
(2001), but point out that the effects due to that dependence on the temperature 
structure of the disk are small and continuous, as discussed by
Jang-Condell \& Sasselov (2004).

Next, we restore irradiation from the central star. The true
midplane temperature is
\begin{equation}
T_{\rm mid}^4 = T_{\rm mid,acc}^4 + T_{\rm irr}^4 \,.
\end{equation}
For details on computing $T_{\rm irr}$, see Sasselov \& Lecar (2000)
and Jang-Condell \& Sasselov (2004). For the central star we used model
parameters for stars of $1~M_{\odot}$ with ages of 1 and 2~Myrs from the
models of Siess et al.~(2000). The models are with a mild overshoot parameter.
We note that models of such young stellar objects are notoriously uncertain.
The span of ages that we consider provides a wide range
of stellar irradiation fluxes and hopefully covers some of this uncertainty.

\section{The Ice Condensation/Sublimation Temperature}

Although the commonly followed rule of thumb for computing the position of
the snow line is simply to take it where the gas temperature drops
to $170$ K (see, e.g. Sasselov \& Lecar 2000), this procedure is too naive. 
As pointed out by Podolak and
Zucker (2004), ice grains will be unstable whenever the grain temperature is
high enough that the rate of water vapor sublimation from the grain
exceeds the rate of water vapor condensation from the surrounding gas. The
grain temperature, in turn, is determined by balancing the
relevant heating and cooling processes. For the case of ice grains in a gas
disk, the grain is heated by the ambient radiation field and
by the release of latent heat when water vapor condenses on the surface. The
grain is cooled by re-radiation and by the removal of latent heat when
ice sublimates. Gas and grains also exchange energy by gas-grain collisions.
The details of the model have been presented elsewhere
(Mekler and Podolak 1994; Podolak and Mekler 1997). In all the
calculations presented in this section,
we assume a fixed gas temperature, $T_{gas}$, and calculate the resulting
grain temperature, $T_{grain}$.

We consider grains in the optically thick midplane of the disk.
The radiative heating flux (energy absorbed per unit area of the grain)
is given by
\begin{equation}
E_{rad,h}=\pi \int_0^\infty Q_{abs}B_{\lambda}(T_{gas})\,d\lambda \,,
\end{equation}
while the radiative cooling flux is given by 
\begin{equation}
E_{rad,c}=\pi \int_0^\infty Q_{emis}B_{\lambda}(T_{grain})\,d\lambda  \,.
\end{equation}
Here $B_{\lambda}(T)$ is the Planck function,
and $Q_{abs}$ and $Q_{emis}$ are the efficiency factors for absorption
and emission of radiation.
We compute $Q_{abs} = Q_{emis}$ from Mie theory; values
depend on grain size and the complex refractive index of the
constituent material.
In this model we considered mixtures of ice and some generic  absorbing
material.  The complex refractive index for ice was  taken from the work of
Warren (1984).
Since ice  is essentially transparent in the visible, where there is
a  peak in the solar spectrum, the temperatures of pure ice grains  exposed
to solar heating can be substantially different from  grains with a small
admixture of material that absorbs in the  visible.  As shown in Podolak and
Mekler (1997), the results  are not sensitive to the details or amount of
absorbing  material provided it produces some absorption in the visible.
For grains in the midplane, where the optical depth to the sun  is high, the
difference in temperature between pure and dirty  ice grains is negligible.

To compute the heating by water vapor condensation, we assume that every
molecule of water vapor that hits the grain condenses and releases a latent
heat of $q$. If $n_{H_2O}$ is the number
density of water molecules and $m_{H_2O}$ is the mass of a water molecule,
the energy flux into the grain due to water condensation is 
\begin{equation}
E_{cond,h}=\frac{n_{H_2O}}2q\sqrt{\frac{2kT_{gas}}{\pi m_{H_2O}}} 
\end{equation}
where $k$ is Boltzmann's constant. We assume that the number density of
water molecules never exceeds the number density for saturation at the ambient
gas temperature or the solar ratio to H$_2$, whichever is lower.

The evaporative cooling is given by 
\begin{equation}
E_{evap,c}=q\frac{P_{vap}(T_{grain})}{\sqrt{\pi m_{H_2O}kT_{grain}}} 
\end{equation}
where $P_{vap}$ is the vapor pressure over ice at the grain temperature.

Finally, the heat flux into the grain from the ambient gas is given by 
\begin{equation}
E_{gas,h}=\frac{n_{H_2}}2\sqrt{\frac{2kT_{gas}}{\pi m_{H_2}}}\frac{jk\left(
T_{gas}-T_{grain}\right) }2 
\end{equation}
where $n_{H_2}$ is the number density of hydrogen molecules, and $j$ is
the number of molecular degrees of freedom ($j=5$ for H$_2$). We
assume a value for $T_{gas}$, equate the total
heating and cooling rates, and solve for $T_{grain}$.
The condition that the grain be stable against evaporation is
that $E_{cond,h}\geq E_{evap,c}$.

Fig. 1 shows the temperature of pure ice grains as a function of the
ambient gas density for $T_{gas} = 150$ K and 170 K.
The solid curves are for
grains of 10 $\mu$m radius and the dashed curves are for grains of 0.1 $\mu$m
radius. While grains in the 150 K gas are all at nearly the same
temperature independent of the gas density, the grains in the 170 K gas have
a temperature that varies both with gas density and grain size.
To explain these results, we first note that if $E_{cond,h} = E_{gas,h} = 0$,
$T_{grain} < T_{gas}$ due to $E_{evap,c}$.
At $T_{gas} = 170$ K,
the rise in $T_{grain}$ with gas density reflects the increasing
importance of $E_{cond,h}$ and $E_{gas,h}$.
The rise is even more pronounced for 0.1 $\mu$m grains
than for 10 $\mu$m grains, since the optical absorption and emission
efficiencies of the former are lower than those of the latter
by two orders of magnitude; non-radiative terms are especially
important for small grains.
At $T_{gas} = 150$ K, the vapor
pressure of water is so low that evaporative cooling and
condensation heating are never important compared to radiation heating and
cooling, and $T_{grain}$ equilibrates to $T_{gas}$.

Fig.~2 shows the energy fluxes due to condensation and sublimation for 0.1 $%
\mu $m (dotted curve) and 10 $\mu $m (dashed curve) ice grains for a
background gas density of $\rho_{gas} = 5\times 10^{-11}$ g cm$^{-3}$. 
Grains become unstable when the temperature goes above
154 K. Note how this sublimation temperature is insensitive
to grain size.

For lower gas densities, the snow line appears at even lower
temperatures, dropping to 150 K at $\rho_{gas}=2\times 10^{-11}$ and to 145
K at $\rho_{gas}=1\times 10^{-11}$ g cm$^{-3}$. This is shown in Fig.~3,
where the gas temperature at the snow line is shown for different values of
the gas density. These values are insensitive to the size of the grain and
its composition (e.g. pure water ice or ice with an admixture of some other
absorber).
In fact, at the snow line, where the  condensation
heating of a grain is almost exactly balanced by  the evaporative cooling, a
much simpler model is possible if  the optical depth to the star is high.
In this case the  radiative heating and radiative cooling of the grain also
balance, and the temperature is given simply by the condition  that the
saturation vapor pressure equal the local partial  pressure of the water
vapor.  For computing the snow line  temperature in the optically thick
midplane, the difference  between this simple model and the detailed model
is too small to be discernable in the figure.

\section{Results and Conclusions}

By adding a modest amount of accretion---$10^{-8}M_{\odot}/{\rm yr} \approx
10^{-5}M_{\rm Jup}/{\rm yr}$---to our standard model of the MMSN at age 1~Myr, we
move the snow transition outwards to 1.6$-$1.8~AU, beyond the orbit of Mars,
as can be seen in Fig.~4.
Observations suggest, however, that the snow line in our Solar System
was located even further out,
near the outer asteroid belt at 2.7 AU, where C-class asteroids
contain some water, albeit a factor of four less than
comets at 5~AU (see the Introduction).
In Table 1, we document the ways in which we can further
increase disk midplane temperatures so as to push the snow line outwards.

We can increase the temperature of the disk by increasing the accretion rate,
but the disk temperature varies only the as 4-th root of $\dot{M}$, while the
lifetime of the disk varies as $1/\dot{M}$. In other words, increasing the
temperature by 10\% comes at the cost of
decreasing the lifetime of the disk by 40\%.
The mass of the MMSN is about 2\% of a solar mass. Therefore, the lifetime
of the MMSN disk with an accretion rate of $10^{-8}M_{\odot}/{\rm yr}$ is
$t_{\rm disk}= M_{\rm disk}/\dot{M} = 0.02~M_{\odot}/10^{-8}M_{\odot}/{\rm yr}
= 2{\times}10^6$ years.

Increasing the surface density of the disk is another
possibility, though more problematic. The
optical depth increases linearly with the surface density, but
the mid-plane temperature scales only as the 4-th root of the optical
depth. And, higher densities are accompanied by higher pressures
which demand higher sublimation temperatures. For a fixed accretion
rate ($10^{-8}M_{\odot}/yr$), varying the disk surface density from 0.1 to 10
times that of the MMSN moves the snow line from 1.6 to only 2.1~AU (see Figure 4).
Also, one gains only slightly from using a flatter density profile (say
$\Sigma \propto r^{-1}$). It is worth noting that Kuchner (2004) derived a 
$\propto r^{-2}$ disk for the 'minimum mass extrasolar nebula'.

Perhaps the most natural resolution to the problem is to boost
the opacity in the disk.
The Rosseland mean opacities increase tenfold for a tenfold decrease in
the maximum grain radius;
the same effect is accomplished by increasing the power-law
exponent of the dust size distribution from 3.0 to 4.0 (e.g., Table 1 of
D'Alessio, Calvet \& Hartmann 2001).
However, such change of grain size properties
might be difficult to justify given recent observations of 1--2~Myr old disks
(Rodmann et al. 2005). The range of possible values of $\kappa$
should be explored further.

In summary, accounting for an accretion rate of $\dot{M} = 10^{-8} M_{\odot}/{\rm yr}$ 
in our standard MMSN disk succeeds in moving the snow line
past Mars. However, moving it out past
3~AU requires, for example, that we simultaneously
increase $\dot{M}$, ${\Sigma}_0$, and $\kappa$ by factors of 2--5
above our assumed baseline values.

\acknowledgements{
M.P. and D.S. acknowledge a grant from the Binational Science Foundation in support of
this work. We thank the referee for a careful reading of our paper and very helpful 
suggestions.}

\clearpage


\begin{table}[!h]
\begin{center}
\begin{tabular}{|c|c|c|c|c|c|c|c|c|}
\hline
$\dot{M}$    &$0.1\times$MMSN && MMSN    && MMSN (high-$\kappa$)     && $10\times$MMSN\\
\hline
            & r(AU)     & T(K) & r(AU)     & T(K)  & r(AU)     & T(K) & r(AU)     & T(K)\\
\hline\hline
$1\times 10^{-8}$  & 1.6 & 151 & 1.7  & 174 & 2.20 & 172 & 2.1 & 185\\
$2\times 10^{-8}$  & 1.8 & 150 & 2.0  & 172 & 2.55 & 170 & 2.5 & 172\\
$4\times 10^{-8}$  & 2.0 & 150 & 2.3  & 170 & 3.00 & 167 & 3.1 & 168\\
$8\times 10^{-8}$  & 2.2 & 149 & 2.7  & 166 & 3.45 & 162 & 3.5 & 174\\
\hline
\end{tabular}
\end{center}
\caption[]{Snow line locations and disk midplane temperatures at 1~Myr age for
four mass accretion rates (high-$\kappa$ stands for a 5-fold opacity increase).}
\end{table}

\clearpage

\begin{figure}
\plotone{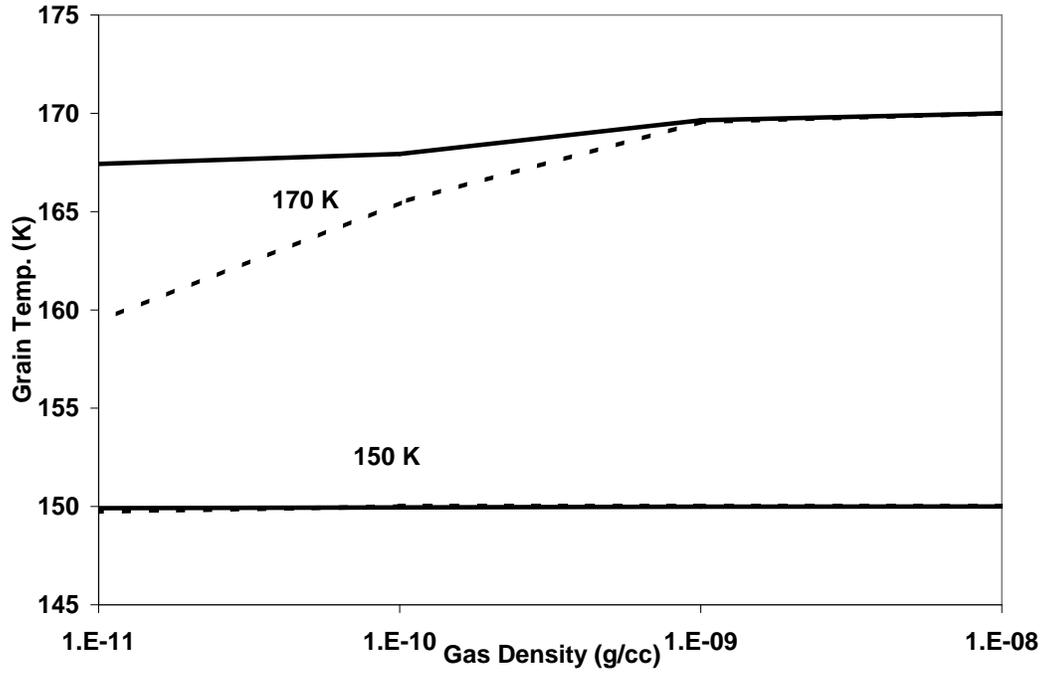}
\caption{Grain temperature as a function of gas density for gas temperatures 
of 150 and 170K. Solid curves are for 10 $\mu$m pure ice grains, dotted 
curves are for 0.1 $\mu$m pure ice
 grains.}
\label{fig1}
\end{figure}

\begin{figure}
\plotone{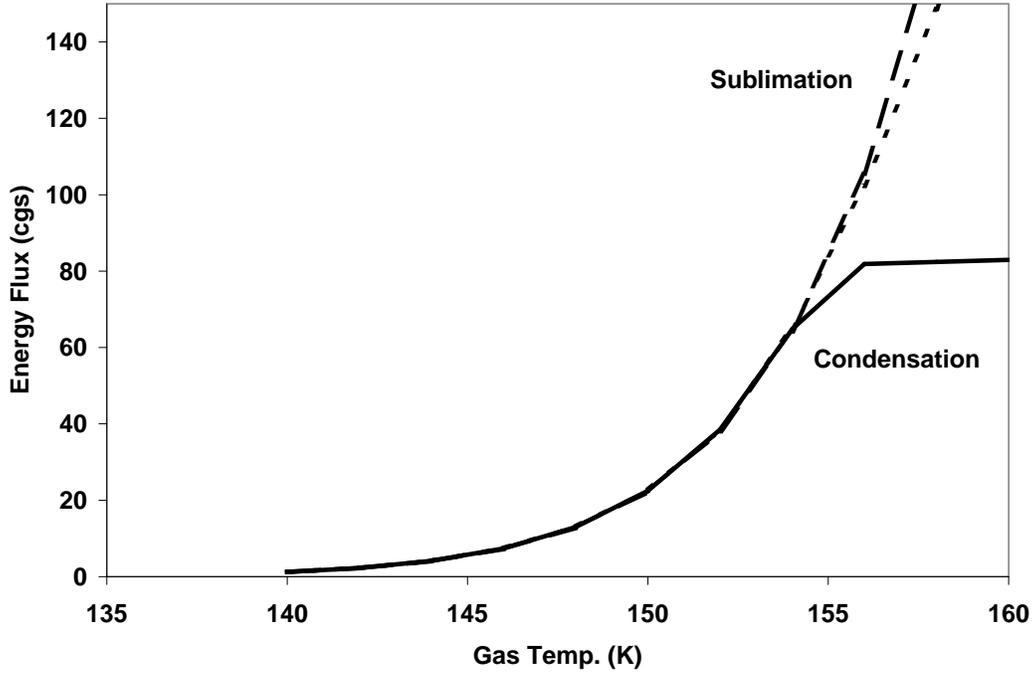}
\caption{Energy fluxes due to condensation (solid) and sublimation for
10 $\mu $m (dashed) and for 0.1 $\mu $m (dotted) pure ice grains. The background
gas density is fixed at $5 \times 10^{-11}$ g cm$^{-3}$.}
\label{fig2}
\end{figure}

\begin{figure}
\plotone{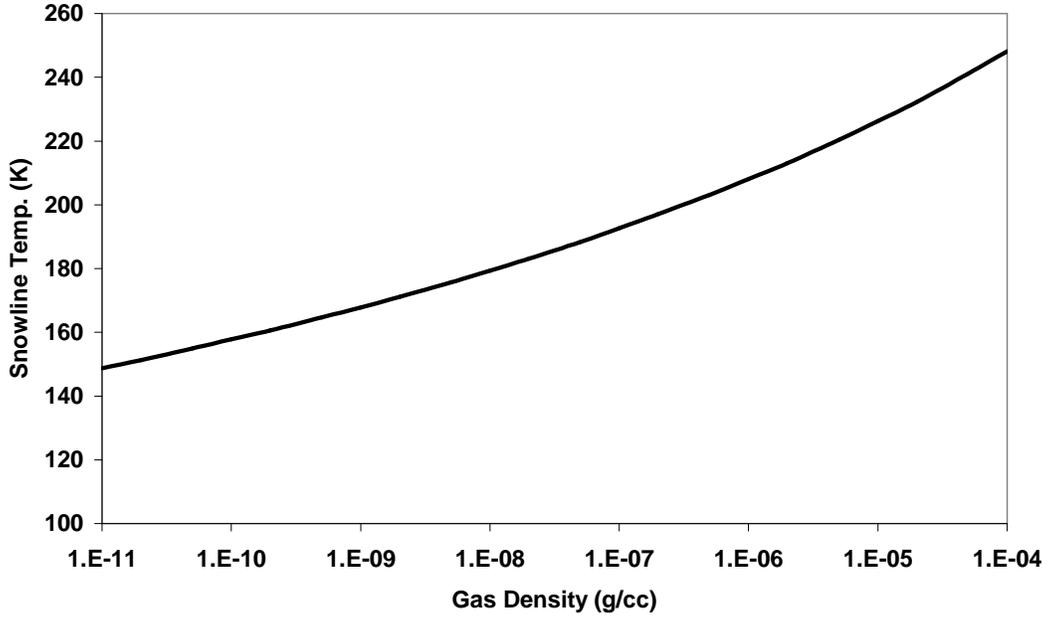}
\caption{Gas temperature at the snow line as a function of gas density.
The result is insensitive to grain size and composition.}
\label{fig3}
\end{figure}

\begin{figure}[!h]
\plotone{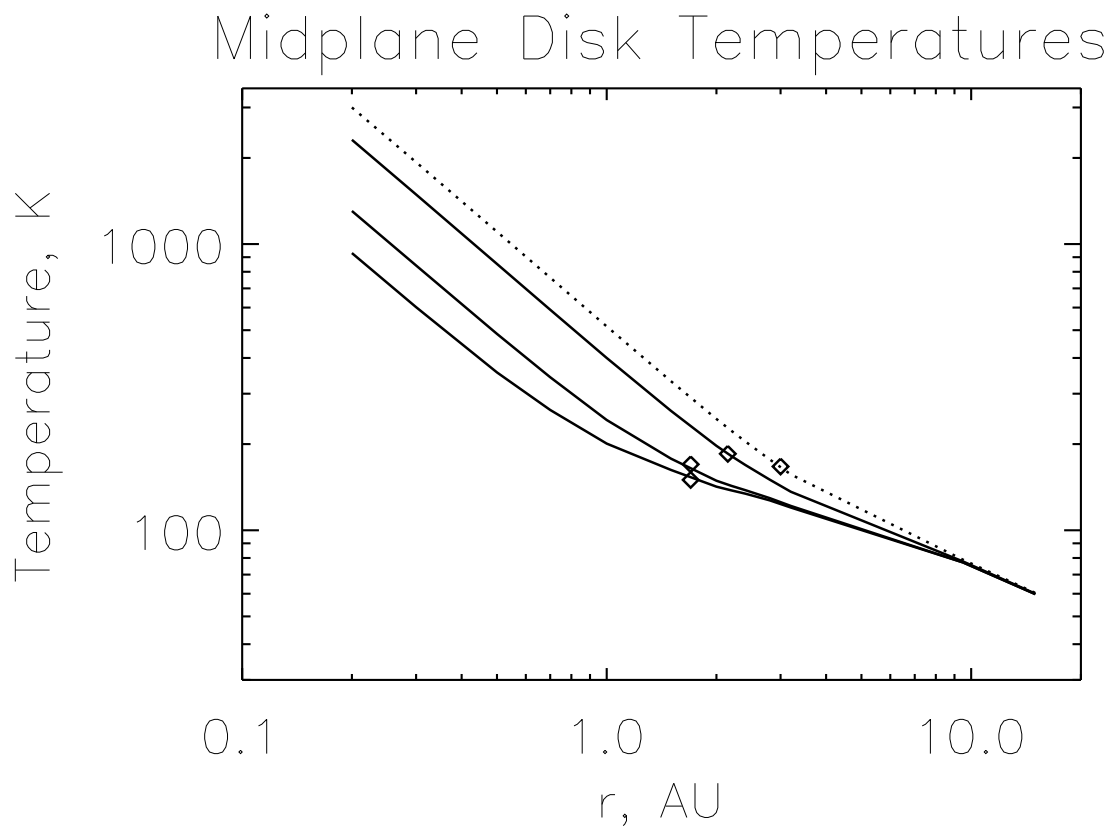}
\caption{Midplane temperatures for disks with masses of 0.1, 1, and 10-times the
MMSN and the snow line locations indicated for each of them (solid lines). 
The disks have a steady
$\dot{M}=10^{-8}M_{\odot}/yr$, ${\Sigma}(r)\propto r^{-3/2}$, and their central stars
are 1~Myr old. Also shown (dotted line) is a MMSN disk model in which the opacity has
been boosted 5-fold and $\dot{M}=4\times 10^{-8}M_{\odot}/yr$. The temperature
gradient becomes shallow where the snow transition occurs because viscous and
irradiation heating exchange dominance at those radii for disk models considered
here.}
\label{fig4}
\end{figure}

\end{document}